 \documentclass[aps,manuscript]{revtex4-1}
 \usepackage{graphicx}
\pdfoutput=1

\begin{document}

 \def\BE{\begin{equation}}
 \def\EE{\end{equation}}
 \def\BA{\begin{array}}
 \def\EA{\end{array}}
 \def\DS{\displaystyle}
 \def\BEA{\begin{eqnarray}}
 \def\EEA{\end{eqnarray}}
 \def\vp{\vec\rho}
 \def\a{\alpha}
 \def\b{\beta}
 \def\l{\lambda}
 \def\vq{\vec q}
 \def\Om{\Omega}
 \def\om{\omega}
 \def\sinc{{\,\rm sinc\,}}
 \def\k{\kappa}
 \def\r{\vec\rho}

\title{Double pass quantum volume hologram}

 \author{Denis~V.~Vasilyev, Ivan~V.~Sokolov}
 \address{ V.~A.~Fock Physics Institute, St.~Petersburg University,
 198504 Petrodvorets, St.~Petersburg, Russia}

 \begin{abstract}
We propose a new scheme for parallel spatially multimode quantum memory for
light. The scheme is based on the propagating in different directions quantum
signal wave and strong classical reference wave, like in a classical volume
hologram and the previously proposed quantum volume hologram \cite{Vasilyev10}.
The medium for the hologram consists of a spatially extended ensemble of cold
spin-polarized atoms. In absence of the collective spin rotation during the
interaction, two passes of light for both storage and retrieval are required,
and therefore the present scheme can be called a double pass quantum volume
hologram. The scheme is less sensitive to diffraction and therefore is capable
of achieving higher density of storage of spatial modes as compared to the thin
quantum hologram of \cite{Vasilyev08}, which also requires two passes of light
for both storage and retrieval. On the other hand, the present scheme allows to
achieve a good memory performance with a lower optical depth of the atomic
sample as compared to the quantum volume hologram. A quantum hologram capable
of storing entangled images can become an important ingredient in quantum
information processing and quantum imaging.
 \end{abstract}

 \maketitle
 \section{Introduction}

A number  of quantum information protocols such as quantum repeaters,
distributed quantum computation, quantum networks etc. require or will greatly
benefit from using a quantum memory. A variety of approaches for storage in
atomic ensembles were developed recently, including the schemes based on
quantum nondemolition (QND) interaction, electromagnetically induced
transparency (EIT), the Raman scattering, and photon echo. A comprehensive
recent review on quantum interfaces between light and matter can be found in
\cite{Hammerer10}. The problem of multimode quantum memories is at the center
of current research due to their potential for enhanced storage capacity, which
is necessary for scalable linear-optical quantum computing \cite{Kok07} and
efficient quantum repeaters \cite{Simon07}.

In this paper we present a new scheme for the spatially multimode quantum
memory for light, which we call a double pass quantum volume hologram. Our
proposal, like the previously considered quantum volume hologram
\cite{Vasilyev10}, makes use of the concept of counter-propagating geometry
which stems from a classical volume hologram, one of the corner stones of
modern holography \cite{Denisyuk62}. When a hologram is written by the
counter-propagating or, in general case, propagating in different directions
signal and reference waves, the two sublattices produced by the waves
interfering in the medium are recorded, each of them storing one quadrature of
the signal field. Since both quadratures are stored, there are no virtual and
real images during the readout, in contrast to a classical hologram recorded in
a single pass of the co-propagating waves.

In the counter-propagating geometry there is no need for phase matching between
the signal and the strong reference waves because their relative phase
oscillates in space hundreds or thousands times. This feature significantly
weakens \cite{Vasilyev10} the diffraction limitation on the transverse spatial
density of stored modes as compared to the co-propagating geometry of the thin
quantum hologram of \cite{Vasilyev08}.

A limitation of the quantum volume hologram with rotating spins \cite{Vasilyev10}
and of the closely related Raman memories for light, extensively studied in
\cite{Nunn07,Mishina07,Shurmacz08,Nunn08,Golubeva10}, is due to their single-pass operation. The
state exchange between light and atoms within the propagation length is associated
with a ``self-erasing'' of quantum state of the input light, which is, roughly
speaking, exponential with oscillations in the longitudinal direction. The higher
efficiency of the state exchange is needed, the larger optical depth of the sample
should be provided.

In contrast to \cite{Vasilyev10}, we consider here the ground state atomic
spins in absence of rotation during the interaction. Hence, during one pass of
light the longitudinal (with respect to the signal wave propagation direction)
quadrature component of the collective spin in a given sublattice does not
change its state and, at the same time, is effectively recorded into one of the
signal wave quadratures within all duration of the interaction cycle. This
feature of our present scheme is inherited from the quantum non-demolition
interaction, typical for the thin quantum hologram, and allows for a good
efficiency of the write-in and read-out by a fixed value of the coupling
constant, which is smaller than for the the previously proposed quantum volume
hologram or the Raman memories.

Due to the  non-demolition in part character of the present scheme, two passes
of light are needed both at the write-in and the read-out stages in order to
``erase'' initial quantum state of the light signal and the atomic spins by the
state exchange, similar to the thin quantum hologram.

We evaluate the transverse spatial density of the stored light modes, and
calculate the average quantum fidelity per transverse mode (per pixel) for the
whole storage cycle. For the initial vacuum (non-squeezed) state of collective
spin, the upper limit on the fidelity is given by $F_{av} \approx 0.845$, and
the fidelity can be made close to 1 given an effective initial (before the
write-in stage) squeezing of first longitudinal modes of the collective atomic
spin.

\section{Single pass operation of volume hologram}

Consider an ensemble of motionless atoms with spin $1/2$ both  in the ground
and in the excited state, located at random positions. The long-lived ground
state spin ${\vec J}^a$ of an atom  is initially oriented in the vertical
direction $x$. A classical off-resonant $x$-polarized plane wave at frequency
$\om_0$ with a slowly varying amplitude $A_x$ (assumed to be real) propagates
in the direction $\vec k_c$, where the corresponding wave vector $\vec k_c$
lies in the  ($y$, $z$) plane. In this paper we investigate the evolution of an
input signal which is represented by a weak quantized $y$-polarized field at
the same frequency $\omega_0$, propagating in $+z$ direction. In what follows
we consider this spatially multimode input field with a slowly varying
amplitude $A_y(\vec r,t)\ll A_x$ in the paraxial approximation. In our scheme
the resulting field strength oscillates in the ($x$, $y$) plane.
\begin{figure}
\centering
\includegraphics[width=65mm]{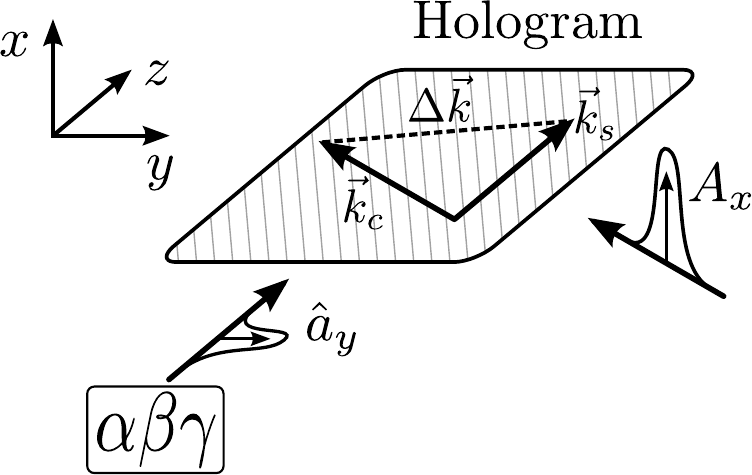}
\caption{Illumination of the hologram  by the write--in (single pass operation).}
 \label{fig1_Scheme}
\end{figure}
Let us introduce the difference wave vector $\Delta\vec k = \vec k_s - \vec
k_c$ with the only non-zero projections $\Delta k_y$ and $\Delta k_z$. One can
consider a thin (that is, of the width much less than $\lambda$) atomic layer
orthogonal to $\Delta\vec k$, where the phase difference between the signal and
the driving classical wave is constant. The quantum non-demolition (QND)
light-matter interaction in each layer leads to two basic effects: (i) the
Faraday rotation of light polarization due to quantum $z$-component of
collective atomic spin of the slice; and (ii) the atomic spin rotation, caused
by the unequal light shifts of the ground state sub-levels with $m_z=\pm 1/2$
in the presence of quantum fluctuations of circular  light polarizations. The
interaction within the slice is described by the well known  QND Hamiltonian
\cite{Hammerer10}:
 \BE
H = \frac{2\pi k_0 |d|^2}{\om_{eg}-\om_0}\int_V d\vec r \sum_a
J_z^a(t) S_z(\vec r,t)\delta(\vec r -\vec r_a).
 \label{interaction}
 \EE
Here $\om_{eg}$ is the frequency of the atomic transition,  $d$ is the dipole matrix
element, and $k_0 =\om_0 c$. For the non-copropagating  signal and driving waves,
the $z$-component of the Stokes vector  $S_z(\vec r,t) = 2A_x\,{\rm Im}\left[A_y(\vec
r,t)\exp(i\Delta\vec k\vec r)\right]$ in (\ref{interaction}) slowly varies within the
slice and rapidly oscillates (i.e. on the scale $1/\Delta k$) along the $\Delta\vec k$
direction. The slowly varying amplitude $A_y$ of the quantized signal field is defined
via
 $$
 A_y(z,\r,t)=\int\frac{dk_z}{2\pi}\int\frac{d\vec
q}{(2\pi)^2}\sqrt{\omega(k)/k_0}\,a_y(\vec k)
 \exp[i\{\vq\r+(k_z-k_0)z-(\omega(k)-\omega_0)t\}],
 $$
where $\vec r=(\r,z)$, $\vec k=(\vq,k_z)$. Here $a_y(\vec k)$ and
$a_y^\dagger(\vec k)$ are the annihilation  and creation operators for the wave
$\vec k$, which obey standard commutation relations $ [a_y(\vec
k),a_y^\dagger(\vec k\,')]=(2\pi)^3\delta(\vec k-\vec k\,')$, $[a_y(\vec
k),a_y(\vec k\,')]=0$. By using these commutation relations in the paraxial
approximation, one finds \cite{Kolobov99} the commutation relation for the
slowly varying amplitude of the quantized signal field $A_y$,
 \BE
 [A_y(z,\r,t),A_y^\dag(z',\r\,',t)] \equiv c\tilde{\delta}(\vec r -
\vec r\,')\approx
 \label{Ay_commutator}
 \EE
 $$
 c\left(1-\frac{i}{k_0}\frac{\partial}{\partial
 z}-\frac{1}{2k^2_0}\nabla^{\,2}_{\bot}\right)\delta(\vec r - \vec r\,').
 $$
We do not consider here quantized waves of other than $\vec k_s$ mean
propagation directions because their evolution is independent of the signal
wave under consideration.

We introduce the density of the collective spin  as $\vec J(\vec r) = \sum_a
\vec J^a \delta(\vec r-\vec r_a)$. The averaged over random positions of the
atoms commutation relation for the $y$, $z$ components of the collective spin
is
 $$
\overline{[J_y(\vec r),J_z(\vec r\,')]} = i\sum_a \langle J_x^a
\rangle \overline{\delta(\vec r-\vec r_a)\delta(\vec r\,'-\vec
r_a)}^a = i n_a \langle J_x^a \rangle \delta(\vec r-\vec r\,').
 $$
Here $n_a$ is the average density of atoms. The field-like canonical variables
for the spin subsystem,
 $$
X(\vec r,t)= J_y(\vec r,t)/\sqrt{n_a \langle J_x^a \rangle},\quad
P(\vec r,t)= J_z(\vec r,t)/\sqrt{n_a \langle J_x^a \rangle},
 $$
obey the canonical commutation relation:
 \BE
 \overline{[X(\vec r,t),P(\vec r\,',t)]} =  i\delta(\vec r-\vec r\,').
 \label{atomic_commutator}
 \EE
The full Hamiltonian of our model includes the energy of free electromagnetic
field and the effective Hamiltonian of QND interaction. The Hamiltonian reads,
 \BE
 H = \int_V d\vec r\left\{ \frac{\hbar
 \omega_0}{c}A_y^\dag(z,\vp,t)\,A_y(z,\vp,t)\, - \right.
 \label{full_hamiltonian}
 \EE
 $$
\left. i\frac{2\pi k_0 |d|^2}{\om_{eg}-\om_0} \sqrt{n\langle J_x^a\rangle}A_x(z,t)
P(z,\vp,t)\big[A_y(z,\vp,t)e^{i\Delta\vec k\vec r}-h.c.\big]\right\}.
 $$
We describe the evolution of our system in the Heisenberg picture.
With the use of commutation relations (\ref{Ay_commutator}) and
(\ref{atomic_commutator}) for the field and atomic variables, after
simple transformations we obtain:
 \BE
 \left(\frac{\partial}{\partial z}-\frac{i}{2k_0}\nabla^{\,2}_{\bot}
+\frac{1}{c}\frac{\partial}{\partial t}\right) A_y(z,\vp,t) =
\frac{\tilde\k}{\sqrt{LT}}\,P(z,\vp,t)e^{-i\Delta\vec k\vec r},
 \EE
 \BE
 \frac{\partial}{\partial t} X(z,\vp,t) =
 \frac{2\tilde\k}{\sqrt{LT}}{\rm Im}\left[A_y(z,\vp,t)e^{i\Delta\vec k\vec r}\right],
 \qquad
 \frac{\partial}{\partial t} P(z,\vp,t) =0.
 \EE
Here $T$ is the duration of the flat-top pulse of driving field  and $L$ is the
atomic cell length. The dimensionless coupling constant
 \BE
\tilde\k=\frac{2\pi k_0 |d|^2}{\hbar (\om_{eg}-\om_0)}\sqrt{n_a\langle
J_x^a\rangle A_x^2 LT},
 \EE
should be of the order of unity for the memory to work (note that $\tilde\k$ is
$\sqrt2$ times the coupling constant $\kappa$ defined in \cite{Vasilyev10} for
volume hologram with rotating spins). The coupling constant can be written as
${\tilde\kappa}^2=2\alpha_0\eta$, where $\alpha_0$ is the resonant optical depth
and $\eta$ is the probability of spontaneous emission \cite{Cerf07}. Since $\eta\ll
1$ is required in order to neglect the effect of spontaneous emission from the upper
level, the usual condition $\alpha_0 = \lambda^2n_aL/2\pi\gg1$ should be fulfilled.
We introduce the Fourier transform via
 \BE
a(z,\vq,t) = \int d\vp\,A_y(z,\vp,t) e^{-i\vq \, \vp},
 \EE
and similar for atomic variables, and arrive to the set of basic  equations in
the Fourier domain:
 \BE
 \label{a_evolution}
 \left(\frac{\partial}{\partial z}+
 i\frac{\vq\,^2}{2k_0}+\frac{1}{c}\frac{\partial}{\partial t}\right)
 a(z,\vq,t) =
 \frac{\tilde\k}{\sqrt{LT}}\,p(z,\vq-\Delta\vec k_y,t)e^{-i\Delta k_z z},
 \EE
 \BE
 \label{xa_evolution}
 \frac{\partial}{\partial t} x(z,\vq,t) = \frac{2\tilde\k}{\sqrt{LT}}{\rm
 Im}\left[a(z,\vq + \Delta\vec k_y,t)e^{i \Delta k_z z}\right],
 \qquad
 \frac{\partial}{\partial t} p(z,\vq,t) =0.
 \EE
Notice that the collective spin quadrature amplitude $p$ responsible for the
Faraday rotation does not evolve, similar to the case of thin quantum hologram
of \cite{Vasilyev08}. Hence, this amplitude is effectively recorded into one of
the signal wave quadratures within all duration $T$ of one interaction cycle.

Let us consider the following boundary condition. We position the center of
atomic sample of length $L$ at $z=0$, and define the input signal wavefront
$A^{(in)}(\r,t)$ and the Fourier amplitudes $a^{(in)}(\vec q,t)$ with respect
to the central ($x$, $y$) plane at $z=0$ in free space. The actual signal field
at the input cell face is related to the defined ``in'' amplitudes as
 \BE
a(-L/2,\vec q,t) = a^{(in)}(\vec q,t)e^{i\frac{q^2}{2k_0}\frac{L}{2}}.
 \label{a_in_def}
 \EE
One can imagine an external lens focusing system which transfers the field
$A^{(in)}(\r,t)$ from its input plane to the plane at $z=0$, so that
$A^{(in)}(\r,t)$ is the input field of the lens system. We arrive to the
following solution of the Eq.~(\ref{a_evolution}):
 \BE
 \label{a_solution}
 a(z,\vq,t) = a^{(in)}(\vq,t)e^{-i\frac{q^2}{2k_0}z}
 + \frac{\tilde\k}{\sqrt{LT}} \int_{-L/2}^z
 dz'p^{(in)}(z',\vq-\Delta\vec k_y)e^{-i\Delta k_z z'}e^{-i\frac{q^2}{2k_0}(z-z')},
 \EE
where $p^{(in)}(z,\vq\,)=p(z,\vq,0)$. We assume pulse duration $T$ to be much
larger than the retardation time at the length $L$ of ensemble, $T\gg L/c$ and
neglect $\sim 1/c$ term in the field evolution equation (\ref{a_evolution}). The
Eq.~(\ref{xa_evolution}) yields,
 \BE
 \label{xa_solution}
 x(z,\vq,t) = x^{(in)}(z,\vq\,)+\frac{2\tilde\k}{\sqrt{LT}}\int_0^t dt'\,{\rm
 Im}\left[a(z,\vq + \Delta\vec k_y,t')e^{i \Delta k_z z}\right].
 \EE
We should have in mind that the field amplitudes are defined as slow in space
in dependence of $z$ coordinate, but the spin amplitudes are not. The Eqs.
(\ref{a_solution}), (\ref{xa_solution}) show that QND interaction with
quantized field $a(z,\vq,t)$ produces fast spatial modulation of collective
spin, and similar fast modulation is read out by the the field. The fast
modulation of collective spin at spatial frequency $\Delta \vec k$ is just a
consequence of the non-collinear geometry of our volume hologram. The volume
hologram is a spatial multi--layer structure with typical spatial period of the
order of $2\pi/\Delta k$. Within this spatial period, the phase difference
between the signal and the driving classical wave changes by $2\pi$ and
therefore changes the type of local circular polarization which is crucial for
QND interaction. This imposes limitations on the atomic motion during storage
time: the atoms should not transport coherence from one layer to another. A
solid state, an optical lattice, or ultra cold atoms should be used.

We go over to slow amplitudes of the collective spin by compensating fast
spatial oscillations present at right side of Eqs.~(\ref{a_solution}),
(\ref{xa_solution}), and perform mode decomposition of the slow amplitudes:
 \BE
 \label{xn_def}
 x_{n}(\vq,t) \equiv \int_{-L/2}^{L/2}dz\,\theta_n(z)
 x(z,\vq-\Delta\vec k_y,t)e^{-i(\Delta k_z-\frac{q^2}{2k_0})z}.
 \EE
The same decomposition is applied to quadrature amplitude $p(z,\vq-\Delta\vec
k_y)$. Here
 \BE
 \theta_n(z) = {\cal N}_n \sqrt{\frac{2}{L}}\, P_n\left(\frac{2z}{L}\right),\qquad
 {\cal N}_n = \sqrt{\frac{2n+1}{2}},
 \EE
$P_n(x)$ is the n-th Legendre polynomial, $n=0,1,2,\dots$, and ${\cal N}_n$ is
the normalization constant. A similar mode decomposition in time domain has
been introduced in \cite{Hammerer06}. Some lower--order amplitudes are given by
 \BEA
 x_{0}(\vq,t) &=& \sqrt{\frac{1}{L}}\int_{-L/2}^{L/2} dz\,
 x(z,\vq-\Delta\vec k_y,t) e^{-i(\Delta k_z-\frac{q^2}{2k_0})z},\\
 x_{1}(\vq,t) &=& 2\left(\frac3{L^3}\right)^{1/2}\int_{-L/2}^{L/2}
 dz\, z\, x(z,\vq-\Delta\vec k_y,t)e^{-i(\Delta k_z-\frac{q^2}{2k_0})z},\\
 x_{2}(\vq,t) &=& 6\left(\frac{5}{L^5}\right)^{1/2}\int_{-L/2}^{L/2} dz
 \left(z^2-\frac{L^2}{12}\right)x(z,\vq-\Delta\vec k_y,t)
 e^{-i(\Delta k_z-\frac{q^2}{2k_0})z}.
 \EEA
We assume that the length $L$ of atomic cell is large as compared to the
wavelength. For large enough angle between the driving and the signal field
wave vectors $\vec k_c$ and $\vec k_s$ (that is, a collinear geometry is not
considered here) one has $\Delta k_z L \gg 1$. Physically this means that our
volume hologram contains many interference layers discussed above.

These new amplitudes allow us to express solutions for atomic and light
variables  in a more compact way. As before, we define the output signal field
amplitude with respect to the central $(x,y)$ plane at $z=0$. The actual signal
field at the output face of the cell is
 \BE
a(L/2,\vec q,t) = a^{(out)}(\vec q,t)e^{-i\frac{q^2}{2k_0}\frac{L}{2}}.
 \label{a_out_def}
 \EE
Here one can imagine an output lens focusing system which in free space
transfers the signal field from the plane $z=0$ to its output plane. Define the
averaged over the pulse duration $T$ signal amplitude as
 \BE
a(\vq\,)=\frac1{\sqrt T}\int_T dt\, a(\vq,t).
 \label{t_averaging}
 \EE
The field propagation equation (\ref{a_solution}) yields,
 \BE
 \label{a_out_solution}
 a^{(out)}(\vq\,) = a^{(in)}(\vq\,) + \tilde\k\, p_{0}^{(in)}(\vq\,).
 \EE
In order to find slow atomic amplitudes  we insert (\ref{a_solution}) into
(\ref{xa_solution}) and make use of the definition (\ref{xn_def}). The
proportional to $e^{-i\Delta k_z z}$ term in (\ref{xa_solution}) after
substitution into (\ref{xn_def}) gives rise to fast spatial oscillations, and
in the limit $\Delta k_z L \gg 1$ vanishes by the integration over $z$. For the
output slow amplitudes of the collective spin, where $x_n^{(out)}(\vq\,) =
x_n(\vq,T)$ and $p_n^{(out)}(\vq\,) = p_n(\vq,T)$, we find
 \BE
 x_{0}^{(out)}(\vq\,) =
 x_{0}^{(in)}(\vq\,)-i\tilde\k
 a^{(in)}(\vq\,) -i\frac{\tilde\k^2}{2}\left[p_{0}^{(in)}(\vq\,)
 - \frac{1}{\sqrt3}p_{1}^{(in)}(\vq\,)\right],
 \label{x0_out}
 \EE
 \BE
 x_{n}^{(out)}(\vq\,) =
 x_{n}^{(in)}(\vq\,)-i\frac{\tilde\k^2}{2}\left[Q_{n,n-1}p_{n-1}^{(in)}(\vq\,)
 + Q_{n,n+1}p_{n+1}^{(in)}(\vq\,)\right],
 \label{xn_out}
 \EE
where $n=1,2,\ldots\,$, and
 \BE
 p_n^{(out)}(\vq\,)=p_n^{(in)}(\vq\,).
 \label{pn_out}
 \EE
Here the matrix $Q$ is given by
 \BE
Q_{n,n-1} = \frac{1}{\sqrt{(2n-1)(2n+1)}},\qquad Q_{n,n+1} = -
\frac{1}{\sqrt{(2n+1)(2n+3)}}.
 \label{Q_matrix}
 \EE
While deriving (\ref{x0_out}, \ref{xn_out}) we made use of the following
property of the Legendre polynomials,
 \BE
\int dx P_n(x) = \frac{1}{2n+1}[P_{n+1}(x) - P_{n-1}(x)].
 \EE
The Eqs.~(\ref{a_out_solution}, \ref{x0_out}) demonstrate that by a single pass
of light, the amplitude $p^{(in)}_{0}(\vq\,)$ of the collective spin is read--out
by the signal, and the amplitude $a^{(in)}(\vq\,)$ of signal field is mapped onto
the atomic amplitude $x_{0}^{(out)}(\vq\,)$. One can also observe in the
Eq.~(\ref{x0_out}) effects of the transfer of the collective spin coherence
between atomic layers in the positive $z$--direction, which are absent in the
QND based thin hologram. The ``active'' spin amplitude $p$ produces a
contribution to the quantized field $a$, which is written into the ``passive''
amplitude $x$ somewhere in the next layers.

By contrast to the QND based scheme of \cite{Vasilyev08}, both quadrature
amplitudes of the light field are stored in a single cycle of light--matter
interaction. This feature of the present scheme is related to its
non--collinear geometry. One can imagine two shifted by a fraction of
wavelength sub--lattices of the collective spin, so that any of two quadrature
amplitudes of the signal is independently stored in the relevant sublattice.

In some sense a single--pass storage and a single--pass retrieval of the signal
light state is achieved in the present scheme. After the first (the write--in)
cycle of interaction one has to rotate the spins by $\pi/2$ relative the
$x$--axis, so that the initial condition for the second (the read--out) cycle
is given by
 \BE
p_{n}^{R(in)}(\vq\,) = x_{n}^{W(out)}(\vq\,),
 \EE
and let another light pulse to read--out the stored state. By applying twice
the {\sl in--out} transformations (\ref{a_out_solution}, \ref{x0_out}), we find
for the output light amplitude:
 \BE
 a^{R(out)}(\vq\,) =  -i\tilde\k^2  a^{W(in)}(\vq\,) + a^{R(in)}(\vq\,)
 + \tilde\k x_{0}^{W(in)}(\vq\,)
 -i\frac{\tilde\k^3}{2}\left[p_{0}^{W(in)}(\vq\,) -
 \frac{1}{\sqrt3}p_{1}^{W(in)}(\vq\,)\right].
 \EE
What we have achieved so far is a classical volume hologram.  Given
$\tilde\k=1$, this result yields the initial quantized signal field
$a^{W(in)}(\vq\,)$ restored with proper amplitude, but quantum fluctuations of
initial amplitudes of the atoms and the read--out light degrade the fidelity of
memory below classical limit. The problem comes from the fact that the light
and matter subsystems do not ``forget'' their initial states during the
interaction.

\section{Double pass quantum volume hologram}

The double pass operation of our volume hologram looks similar to the case of
the QND based thin hologram. In order to compensate initial noise terms and to
go beyond the classical limit, we introduce at the writing stage the second
pass through the atomic sample of both (the signal and the classical) light
waves in the same directions $\vec k_s$ and $\vec k_c$. After the first pass
one has to rotate the spins around the $x$-axis by $\pi/2$ pulse of auxiliary
magnetic field, and to apply phase shift of $\pi/2$ to the output signal wave
relative the classical driving wave within the interaction volume. The
transformation of light and matter variables between the first and the second
pass, which is described by the same equations as before, is given by
 \BE
 \label{phase_shift}
 a^{W(2,in)}(\vq\,) = ia^{W(1,out)}(\vq\,), \quad
 x_{n}^{W(2,in)}(\vq\,) = -p_{n}^{W(1,out)}(\vq\,), \quad
 p_{n}^{W(2,in)}(\vq\,) = x_{n}^{W(1,out)}(\vq\,).
 \EE
After all three steps of the write-in procedure we arrive to the following {\sl
in--out} relations for the double--pass operation:
 \BE
 \label{light_two_passes}
 a^{W(out)}(\vq\,) = i(1-\tilde\k^2)a^{W(in)}(\vq\,) + i\tilde\k
 \left(1-\frac{\tilde\k^2}{2}\right)p_{0}^{W(in)}(\vq\,) +
 \tilde\k x_{0}^{W(in)}(\vq\,) +
 i\frac{\tilde\k^3}{2\sqrt3}p_{1}^{W(in)}(\vq\,),
 \EE
 $$
  x_{0}^{W(out)}(\vq\,) = \tilde\k\left(1-\frac{\tilde\k^2}2\right)
 a^{W(in)}(\vq\,)-
 \left(1-\tilde\k^2+\frac{\tilde\k^4}{6}\right)p_{0}^{W(in)}(\vq\,)
 -i\frac{\tilde\k^2}{2}\left[x_{0}^{W(in)}-\frac{1}{\sqrt3}x_{1}^{W(in)}\right] +
  $$
 \BE
 \label{xa_two_passes}
   \frac{\tilde\k^4}{4}\left[\frac{1}{\sqrt3}p_{1}^{W(in)}(\vq\,)
 - \frac{1}{3\sqrt5}p_{2}^{W(in)}(\vq\,)\right],
 \EE
and
 \BE
 p_{0}^{W(out)}(\vq\,) =
 x_{0}^{W(in)}(\vq\,)-i\tilde\k
 a^{W(in)}(\vq\,) -i\frac{\tilde\k^2}{2}\left[p_{0}^{(in)}(\vq\,)
 - \frac{1}{\sqrt3}p_{1}^{(in)}(\vq\,)\right],
 \label{p0_two_passes}
 \EE
 \BE
 p_{1}^{W(out)}(\vq\,) =
 x_{1}^{W(in)}(\vq\,)-i\frac{\tilde\k^2}{2}\left[\frac{1}{\sqrt{3}}p_0^{W(in)}(\vq\,)
 -\frac{1}{\sqrt{15}}p_2^{W(in)}(\vq\,)\right].
 \label{pn_two_passes}
 \EE
The double pass read--out procedure looks similar and is described by the Eqs.
(\ref{light_two_passes}--\ref{pn_two_passes}), where one has to substitute the
label W to R.

We have omitted expressions for higher--order spin amplitudes because the
output amplitude of the signal wave by the read--out $a^{R(out)}(\vq\,)$ is
decoupled from these observables, as seen from the
Eq.~(\ref{light_two_passes}). The result (\ref{light_two_passes}) also shows
that both by the double pass write--in and the read--out, the output signal
wave amplitude is completely independent of the input wave if the coupling
constant is chosen to be $\tilde\k=1$, which is one of conditions for efficient
state exchange between light and matter. In this sense our present model works
as the thin quantum hologram of \cite{Vasilyev08}, where by the first pass the
input signal amplitude is imprinted into the collective spin wave. After the
transformation given by the Eq. (\ref{phase_shift}) and the second pass, the
relevant collective spin amplitude effectively  cancels the input signal
contribution to the output quantized field.

At the same time, the initial state of light is stored in the collective spin
amplitudes $x_{0}^{W(out)}$ and $p_{0}^{W(out)}$. As a result of the atom-light
interaction during two passes with properly adjusted coupling constant we have
the write stage completed. Atoms and light have exchanged their initial quantum
states.

In order to read out the memory one has to repeat the procedure. The output
quantized field amplitude for the whole write--in and read--out cycle is found
by using the {\sl W(out)} amplitudes as the {\sl R(in)} variables. Finally we
arrive to
 $$
 a^{R(out)}(\vq\,) = \tilde\k^2(2-\tilde\k^2)a^{W(in)}(\vq\,) +
 i(1-\tilde\k^2)a^{R(in)}(\vq\,) -
 \tilde\k\left(1-\frac{3\tilde\k^2}{2}+\frac{\tilde\k^4}{3}\right)p_{0}^{W(in)}(\vq\,)+
 $$
 \BE
 \label{light_read_out}
  i\tilde\k(1-\tilde\k^2)x_{0}^{W(in)}(\vq\,)+
 i\frac{\tilde\k^3}{\sqrt3}x_{1}^{W(in)}(\vq\,)-
 \frac{\tilde\k^3}{2\sqrt3}(1-\tilde\k^2)p_{1}^{W(in)}(\vq\,)-
 \frac{\tilde\k^5}{6\sqrt5}p_{2}^{W(in)}(\vq\,).
 \EE
 %
%
%
%
%
One optimizes the memory performance by choosing the coupling constant
$\tilde\k=1$. The input and the output variables of the total write--readout
protocol of the volume quantum hologram are related as
 \BE
 a^{R(out)}(\vq\,) = a^{W(in)}(\vq\,) + f(\vq\,),
 \EE
where $f(\vq\,)$ is the added noise term. This transformation is  analogous to
the one describing quantum holographic teleportation of an optical image
\cite{Sokolov01,Gatti04} and the thin quantum hologram of \cite{Vasilyev08}.
The noise contributions specific for our model of memory is given by
 \BE
 \label{Noise}
 f(\vq\,)=\frac{i}{\sqrt3}x_1^{W(in)}(\vq\,)
 +\frac{1}{6} p_0^{W(in)}(\vq\,)-\frac{1}{6\sqrt5}p_2^{W(in)}(\vq\,),
 \EE
where we assume $\tilde\k = 1$. We perform the backward Fourier transform $\vec
q \to \vec \rho$ and arrive at
 \BE
 A^{R(out)}(\vp\,) = A^{W(in)}(\vp\,) + F(\vp\,), \qquad
 F(\vp\,) \equiv \frac{1}{\sqrt{2}}[F_X(\vec\rho\,) + i F_P(\vec\rho\,)],
 \EE
where we apply the same definition of real (Hermitian) noise quadrature amplitudes
$F_{X,P}( \vec\rho\,)$ as in \cite{Vasilyev08}. Consider orthogonal pixellized spatial
modes associated with the field amplitudes averaged over the surface $S_j$ of
square pixel $j = 1,\ldots,N$ of area $S$. The averaged noise quadrature
amplitudes and the noise covariance matrix are defined via
 \BE
F_{X,P}(j) = \frac{1}{\sqrt{S}}\int_{S_j}d\r\, F_{X,P}(\r\,),
 \EE
 \BE
C^X(i,j)= \langle F_X(i)F_X(j)\rangle, \qquad C^P(i,j) = \langle
F_P(i)F_P(j)\rangle.
 \EE
The collective spin noise quadrature amplitudes look like
 \BE
\frac{1}{\sqrt{2}}F_X(j) = -\frac{1}{\sqrt{3}} {\rm Im} X_1^{W(in)}(j) +
\frac{1}{6} {\rm Re} P_0^{W(in)}(j)-\frac{1}{6\sqrt{5}} {\rm Re}P_2^{W(in)}(j),
 \EE
 \BE
\frac{1}{\sqrt{2}}F_P(j) = \frac{1}{\sqrt{3}} {\rm Re} X_1^{W(in)}(j) +
\frac{1}{6} {\rm Im} P_0^{W(in)}(j) - \frac{1}{6\sqrt{5}} {\rm
Im}P_2^{W(in)}(j).
 \EE
As seen from the Eq.~(\ref{xn_def}),  the real and the imaginary part of the
amplitudes $X_n$ can be related to real quadrature amplitudes $X_{n,c}$ and
$X_{n,s}$ of the  longitudinal collective spin modes whose spatial profile is
given by
 $$
\sqrt{2}\,\theta_n(z)\cos\left[\left(\Delta k_z-\frac{q^2}{2k_0}\right)z\right], \qquad
\sqrt{2}\,\theta_n(z)\sin\left[\left(\Delta k_z-\frac{q^2}{2k_0}\right)z\right].
 $$
We arrive to
 \BE
{\rm Re} X_n(j) = \frac{1}{\sqrt{2}}X_{n,c}(j), \qquad {\rm Im} X_n(j) = -
\frac{1}{\sqrt{2}}X_{n,s}(j),
 \EE
and similar definitions are applied to the amplitudes $P_n$.
In the limit of large number of the interference layers disused above, $|\Delta k_z
L| \gg 1$, the ``$\sin$'' and ``$\cos$'' modes are orthogonal and correspond to
independent degrees of freedom of the spin coherence.

The quality of quantum state transfer $|\psi^{(in)}\rangle \to |\psi^{(out)}\rangle$
is quantified via the fidelity parameter
$F=|\langle\psi^{(in)}|\psi^{(out)}\rangle|^2$.  Assume the input signal field to be
in a spatially multimode coherent state. For the image field in a coherent state,
decomposed over $N$ pixellized modes, the fidelity is given \cite{Gatti04} by
 \BE
F_N = \left[{\rm det}\{\delta_{ij} + C^X(i,j)\}\,{\rm
det}\{\delta_{ij} + C^P(i,j)\}\right]^{-1/2}.
 \label{fidelity_matrix}
 \EE
We evaluate the noise covariance matrix assuming all longitudinal and
transverse modes of the collective atomic spin to be initially in vacuum state,
$\langle \left(X_{n,i}^{W(in)}(j)\right)^2\rangle =
\langle\left(P_{n,i}^{W(in)}(j)\right)^2\rangle = 1/2$, where  $i=c,s$. This
yields,
 \BE
 C^X(i,j) =  C^P(i,j) = \frac{11}{60}\delta_{ij}, \qquad F_N =
 \left(\frac{60}{71}\right)^{N}.
 \EE
As shown in \cite{Gatti04}, the fidelity of quantum state  transfer for simple
multipixel arrays scales approximately as the $N$-th power of the quantity
which is called average fidelity per pixel, $F_{av}=(F_N)^{1/N}$. For the input
signal prepared in a multipixel coherent state we obtain the average fidelity
per pixel $F_{av}\approx0.845$ for the whole write-in and read-out cycle of our
quantum hologram. This is significantly higher than the classical benchmark
$F_{av} = 1/2$ for the holographic storage of an optical image in multipixel
coherent state, and than the limit of $F_{av} = 2/3$ on quantum cloning.

Theoretically, in order to achieve better fidelity one could perform squeezing
of each of the added noise quadrature amplitudes in (\ref{Noise}). Given
perfect squeezing, which we do not consider here, this would lead to maximum
fidelity $F_{av}\to 1$.

The optimal value of the coupling constant $\tilde\kappa = 1$ corresponds to
$\kappa=\sqrt{2}$ for the quantum volume hologram of \cite{Vasilyev10}, which
gives there the storage and retrieval efficiency of $0.10-0.15$. This is
significantly less than the non-zero quantum capacity limit on the efficiency
of $0.5$ and leads, by such a low value of the coupling constant, to
performance on a par with a classical hologram only. It should be noted that the experimentally achievable atomic samples have spatially dependent distribution of atomic density. The required homogeneous coupling constant could be obtained by choosing a proper transversal profile of the classical driving wave.

The resolving power in space of the double--pass quantum hologram can be
characterized by a number of transverse modes (pixels), effectively stored and
retrieved in the memory. Similar to the quantum volume hologram of
\cite{Vasilyev10} and the spatially resolving Raman memories \cite{Golubeva10},
 %
 %
by the write--in and read--out in the same direction the diffraction does not
modify the light--matter interaction and does not impose limitations on the
resolving power which is finally determined by the sample geometry: the
effectively stored orthogonal modes should propagate within the atomic
ensemble. Note that the diffraction by propagation at the cell length can be
compensated by thin lenses, and our definitions (\ref{a_in_def},
\ref{a_out_def}) of the input and output field amplitudes account for this
compensation in explicit form. The estimate of the number of stored modes given
in \cite{Vasilyev10} applies to the scheme considered here as well.

\section{Conclusion}

We have elaborated a novel version of spatially multimode quantum memory for
light -- double pass quantum hologram. Similar to our previous proposals of
thin quantum hologram \cite{Vasilyev08} and  volume quantum hologram
\cite{Vasilyev10}, the present scheme can be considered as an extension of
classical holograms  into the quantum domain.  In our scheme, different spatial
modes of the incoming field are stored in the corresponding orthogonal spatial
modes of the long--lived collective spin of an ensemble of ultra--cold atoms in
absence of spin rotation in external magnetic field. The double pass quantum
volume hologram inherits some features of the thin and the volume quantum
hologram. It is able to store transverse modes of an input light signal with
the same high density as the quantum volume hologram and the Raman memories do,
and it requires a fixed and relatively low optical depth as the thin hologram
does. On the other hand, both the write--in and the read--out cycles of the
double pass quantum hologram require two passes of light, and for a high
fidelity storage (with the average fidelity per pixel exceeding $0.845$),
preparation of the collective spin in a quadrature squeezed state is also
required. Although we considered spin $1/2$ atoms, our analysis can be in
principle applied to alkali atoms provided that the optical detuning
significantly exceeds the excited state hyperfine splitting \cite{Hammerer10}.

This research has been funded by the European Commission FP7 under the grant
agreement n° 221906, project HIDEAS. The authors also acknowledge the support
of the Russian Foundation for Basic Research under the projects 08-02-00771 and
08-02-92504. Part of the research was performed within the framework of GDRE
``Lasers et techniques optiques de l'information''.

 \end{document}